\pdfoutput=1

\documentclass[11pt,twoside,a4paper,cmspaper,final,collab]{cms-tdr}
\def\svnVersion{416587}\def\cmsCernNoTag{CERN-EP/2017-018}\def\cmsCernDate{\today}\def\cmsMessage{Published in the European Physical Journal C as \href{http://dx.doi.org/10.1140/epjc/s10052-017-5030-3}{\doi{10.1140/epjc/s10052-017-5030-3.}}}

\begin{document}\cmsNoteHeader{TOP-15-015}

\hyphenation{had-ron-i-za-tion}
\hyphenation{cal-or-i-me-ter}
\hyphenation{de-vices}
\RCS$Revision: 411530 $
\RCS$HeadURL: svn+ssh://svn.cern.ch/reps/tdr2/papers/TOP-15-015/trunk/TOP-15-015.tex $
\RCS$Id: TOP-15-015.tex 411530 2017-06-20 10:19:44Z alverson $

\newlength\cmsFigWidth
\ifthenelse{\boolean{cms@external}}{\setlength\cmsFigWidth{0.85\columnwidth}}{\setlength\cmsFigWidth{0.4\textwidth}}
\ifthenelse{\boolean{cms@external}}{\providecommand{\cmsLeft}{upper\xspace}}{\providecommand{\cmsLeft}{left\xspace}}
\ifthenelse{\boolean{cms@external}}{\providecommand{\cmsRight}{lower\xspace}}{\providecommand{\cmsRight}{right\xspace}}
\ifthenelse{\boolean{cms@external}}{\providecommand{\cmsTableResize[1]}{\resizebox{\columnwidth}{!}{#1}}}{\providecommand{\cmsTableResize[1]}{\relax{#1}}}
\ifthenelse{\boolean{cms@external}}{\providecommand{\cmsFigPage}{\begin{figure}[htbp]}}{\providecommand{\cmsFigPage}{\begin{figure}[htb]}}
\newcommand{\MADSPIN}{\textsc{Madspin}\xspace}
\newcommand{\ptrel}{\ensuremath{p_{\text{rel,T}}}\xspace}
\newcommand{\mjet}{\ensuremath{m_{\text{jet}}}\xspace}
\newcommand{\mt}{\ensuremath{m_{\PQt}}\xspace}
\newcommand{\mtMC}{\ensuremath{m_{\PQt}}\xspace}
\newcommand{\ptone}{\ensuremath{p_{\mathrm{T,1}}}\xspace}
\newcommand{\pttwo}{\ensuremath{p_{\mathrm{T,2}}}\xspace}
\newcommand{\ptveto}{\ensuremath{p_{\text{T,veto}}}\xspace}
\newcommand{\pb}{\unit{pb}}
\newcommand{\fb}{\unit{fb}}
\newcommand{\powpyth}{\POWHEG{}+\PYTHIA}
\newcommand{\madpyth}{\MADGRAPH{}+\PYTHIA}
\newcommand{\mcherwig}{\MCATNLO{}+\HERWIG}
\newcolumntype{.}{D{.}{.}{3.4}}
\newcolumntype{x}{D{.}{.}{4.4}}

\cmsNoteHeader{TOP-15-015}
\title{Measurement of the jet mass in highly boosted \ttbar events from pp collisions at $\sqrt{s}=8$\TeV}

\date{\today}

\abstract{
The first measurement of the jet mass $m_{\text{jet}}$ of top quark jets
produced in $\ttbar$ events from pp collisions at $\sqrt{s}=8$\TeV
is reported for the jet with the largest transverse momentum \pt
in highly boosted hadronic top quark decays.
The data sample, collected with the CMS detector,
corresponds to an integrated luminosity of 19.7\fbinv.
The measurement is performed in the lepton+jets channel
in which the products of the semileptonic decay
$\PQt \to  \PQb\PW$ with $\PW\to\ell \nu$
where $\ell$ is an electron or muon,
are used to select $\ttbar$ events with large Lorentz boosts.
The products of the fully hadronic decay
$\PQt \to \PQb\PW$ with $\PW\to\PQq\PAQq'$
are reconstructed using a single Cambridge--Aachen jet
with distance parameter $R=1.2$, and $\pt>400$\GeV.
The $\ttbar$ cross section as a function of $m_{\text{jet}}$ is unfolded
at the particle level and is used to test the modelling of
highly boosted top quark production.
The peak position of the $m_{\text{jet}}$ distribution is sensitive to the top
quark mass $m_{\PQt}$, and the data are used to extract a value of $m_{\PQt}$
to assess this sensitivity.
}

\hypersetup{%
pdfauthor={CMS Collaboration},%
pdftitle={Measurement of the jet mass in highly boosted t-tbar events from pp collisions at sqrt(s) = 8 TeV},%
pdfsubject={CMS},%
pdfkeywords={CMS, physics, top quark cross section, top quark mass}}

\maketitle

\section{Introduction}
\label{sec:intro}
The top quark may play a special role in the standard model (SM) of particle physics
owing to its large mass and its possible importance in electroweak symmetry
breaking~\cite{Hill:2002ap, Degrassi:2012ry}.
Measurements of \ttbar production provide crucial
information about the accuracy of the SM near the electroweak
scale~\cite{Buckley:2015nca, Buckley:2015lku}, and in
assessing the predictions of quantum chromodynamics (QCD) at large mass scales.
In turn, they can be used to determine the fundamental parameters of the theory,
such as the strong coupling constant or the top quark
mass~\cite{Chatrchyan:2013haa, Khachatryan:2016mqs}.

{\tolerance=1200
Previous differential measurements of the \ttbar production cross
section~\cite{Abazov:2010js, Chatrchyan:2012saa, CDF:2013gna, Abazov:2014vga, Aad:2014zka, Aad:2015eia,
Khachatryan:2015oqa, Khachatryan:2015fwh, Aaboud:2016iot} at the Fermilab Tevatron and CERN LHC
show excellent agreement with SM predictions.
However, investigations of top quarks with very large transverse momenta $\pt$
have proven to be difficult, since
in this kinematic range the decays of the top quark to fully hadronic final states become
highly collimated and merge into single jets.
In this highly boosted regime, the \ttbar reconstruction efficiency
deteriorates for previous, more-traditional measurements. Special reconstruction
techniques based on jet substructure are often used to improve the
measurements~\cite{Aad:2015hna, Khachatryan:2016gxp} or to implement searches for new
physics~\cite{Chatrchyan:2012ku, Aad:2012dpa, Aad:2012ans, Aad:2013nca,
Chatrchyan:2013lca, Aad:2014xra, Khachatryan:2015axa, Aad:2015fna,
Khachatryan:2015sma, Khachatryan:2015mta, Aad:2015voa}.
A detailed understanding of jet substructure observables,
and especially the jet mass \mjet, is crucial for LHC analyses of highly boosted topologies.
While measurements of \mjet corrected to the particle level
have been carried out for light-quark and gluon jets~\cite{ATLAS:2012am, Chatrchyan:2013vbb},
the \mjet distribution for highly boosted top quarks has not yet been measured.
\par}

Apart from testing the simulation of \mjet in fully hadronic top quark decays,
the location of the peak of the \mjet distribution
is sensitive to the top quark mass \mt~\cite{Hoang:2008xm}.
This measurement therefore provides an alternative method
of determining \mt in the boosted regime, independent of previous mass
measurements~\cite{Aaltonen:2012ra, Aad:2015nba, Aad:2015waa, Khachatryan:2015hba,
Khachatryan:2016wqo, Aaboud:2016igd}.
Calculations from first principles have been
performed in soft collinear effective
theory~\cite{Bauer:2000ew, Bauer:2000yr, Bauer:2001ct, Bauer:2001yt}
for the dijet invariant mass distribution from highly boosted top quark
production in $\Pep\Pem$ collisions~\cite{Fleming:2007qr,Fleming:2007xt},
and work is ongoing to extend this to the LHC environment~\cite{Hoang:2015vua, Butenschoen:2016lpz}.
Such calculations account for perturbative and nonperturbative effects, and
provide particle-level predictions.
Once predictions for the LHC become available,
the measurement of the \mjet distribution can lead to an extraction of \mt
without the ambiguities that arise from the unknown relation between \mt in
a well-defined renormalisation scheme and the top quark mass parameter used in Monte Carlo (MC)
simulations~\cite{Butenschoen:2016lpz, Moch:2014tta, Hoang:2014oea, Corcella:2015kth}.

We present the first measurement of the differential \ttbar production cross section
as a function of the leading-jet mass, where leading refers to the jet with the highest $\pt$.
The measurement is based on data from $\Pp\Pp$ collisions at $\sqrt{s} = 8\TeV$,
recorded by the CMS experiment at the LHC in 2012
and corresponding to an integrated luminosity of 19.7\fbinv.
It is performed on \ttbar events in which the leading jet includes
all $\PQt \to\PQb \PWp \to \PQb \qqbar'$ decay products.
The other top quark is required to decay through the semileptonic mode
$\PAQt \to \PAQb \PWm \to \PAQb \ell \PAGn_\ell$,
where $\ell$ can be either an electron or muon.
The use of charge-conjugate modes is implied throughout this article.
The semileptonic top quark decay serves as a means for selecting \ttbar events without biasing the \mjet
distribution from the fully hadronic top quark decay.
The highly boosted top quark jets used in the measurement are defined through the Cambridge--Aachen (CA)
jet-clustering algorithm~\cite{CACluster1,CACluster2}
with a distance parameter $R=1.2$ and $\pt>400\GeV$.
The \mjet distribution is unfolded to the particle level
and compared to predictions from MC simulations. The measurement is
also normalised to a fiducial-region total cross section defined below,
and shows the expected sensitivity
to the value of \mtMC. An extraction of the value of \mtMC is performed to
assess the overall sensitivity of the measurement.

\section{The CMS detector}
\label{sec:detector}
The central feature of the CMS detector is a superconducting solenoid
of 6\unit{m} internal diameter, providing a magnetic field of
3.8\unit{T}. A silicon pixel and strip tracker,
a lead tungstate crystal electromagnetic
calorimeter (ECAL), and a brass and scintillator hadron calorimeter (HCAL),
each composed of a barrel and two endcap sections
reside within the magnetic volume. In addition
to the barrel and endcap detectors, CMS has extensive forward
calorimetry.
Muons are detected using four layers of gas-ionization detectors embedded
in the steel flux-return yoke of the magnet.
The inner tracker measures charged particle
trajectories within the pseudorapidity range $\abs{\eta} < 2.5$.
A two-stage trigger system~\cite{Khachatryan:2016bia} is used to select for analysis
$\Pp\Pp$ collisions of scientific interest.
A more detailed description of the CMS detector, together with a
definition of the coordinate system and relevant kinematic
variables, can be found in Ref.~\cite{Chatrchyan:2008zzk}.

\section{Event reconstruction}
\label{sec:reco}
The CMS experiment uses a particle-flow (PF) event
reconstruction~\cite{CMS-PAS-PFT-09-001, CMS-PAS-PFT-10-001},
which aggregates input from all subdetectors.
This information includes charged particle tracks from the tracking system and
energies deposited in the ECAL and HCAL,
taking advantage of the granularity of the subsystems.
Particles are classified as electrons, muons, photons, and charged and neutral hadrons.
Primary vertices are reconstructed
using a deterministic annealing filter algorithm~\cite{Chatrchyan:2014fea}.
The vertex with the largest sum in the associated track $\pt^2$ values
is taken to be the primary event vertex.

Muons are detected and measured in the pseudorapidity range $\abs{\eta} < 2.1$
using the information collected in the muon and tracking
detectors~\cite{Chatrchyan:2012xi}. Tracks from muon candidates
must be consistent with a muon originating
from the primary event vertex, and satisfy track-fit quality
requirements~\cite{Chatrchyan:2013sba}.

Electrons are reconstructed in the range $\abs{\eta} < 2.1$,
by combining tracking information with energy deposits in the
ECAL~\cite{Khachatryan:2015hwa, Chatrchyan:2013dga}.
Electron candidates are required to originate from the primary event vertex.
Electrons are identified through the information on the
energy distribution in their shower, the
track quality, the spatial match between the track and
electromagnetic cluster, and the fraction of total cluster energy in the
HCAL. Electron candidates that
are consistent with originating from photon conversions in the
detector material are rejected.

Since the top quark decay products can be collimated at high values of top quark \pt,
no isolation requirements on the leptons are imposed in either the trigger or
in the offline selections (see Section~\ref{sec:datasets}).
The imbalance in event $\vec{\pt}$ is quantified as the missing transverse momentum vector \ptvecmiss,
defined as the projection on the plane perpendicular to the beams of the
negative vector sum of the momenta of all PF candidates in the event.
Its magnitude is referred to as \ptmiss.

The PF candidates are
clustered into jets by using the \FASTJET 3.0 software
package~\cite{FastJet}.
Charged hadrons associated with event vertices other than the primary event vertex
are removed prior to jet clustering. Isolated leptons (either electron or muon) are not
part of the input list for jet finding~\cite{CMS-PAS-PFT-09-001, CMS-PAS-PFT-10-001}.
Small-radius jets are clustered with the anti-\kt
jet-clustering algorithm~\cite{Cacciari:2008gp} with a distance parameter
$R=0.5$ (AK5 jets). These small-radius jets are used at the trigger level, in the
first steps of the event selection, and for the identification of jets
coming from the hadronisation of $\PQb$ quarks.
If a nonisolated lepton candidate
is found within the angular distance ${\Delta R < 0.5}$
of an AK5 jet, its four-momentum is subtracted from that of the jet
to avoid double counting of energy and ensure proper jet energy corrections.
The angular distance is given by $\Delta R = \sqrt{\smash[b]{(\Delta \phi)^2 + (\Delta \eta)^2}}$,
where $\Delta \phi$ and $\Delta \eta$ are the differences in azimuthal angle (in radians) and
pseudorapidity, respectively, between the directions of the lepton and jet.
Large-radius jets are obtained by using the CA
jet-clustering algorithm~\cite{CACluster1,CACluster2} with $R=1.2$ (CA12 jets).
When a lepton candidate is found among the PF candidates clustered into a CA12 jet,
its four-momentum is subtracted from that of the CA12 jet.
In this paper, the unmodified term ''jet'' will refer to the broad CA12 jets.

All jets could contain neutral particles from additional $\Pp\Pp$
collisions in the same or nearby beam crossings (pileup).
This extra contribution is subtracted based on the
average expectation of the pileup in the jet catchment area~\cite{Cacciari:2008gn}.
This is done by calculating a correction for the average offset energy density in each event
as a function of the number of primary vertices~\cite{Chatrchyan:2011ds, Khachatryan:2016kdb}.
The AK5 jets are identified as originating from the fragmentation of a $\PQb$ quark
with the combined secondary vertex algorithm (CSV)~\cite{Chatrchyan:2012jua}.
A tight operating point is used,
which has a misidentification probability of 0.1\%
for tagging light-parton jets with an average \pt of about 80\GeV,
and an efficiency of about 50\% for a heavy-flavour
jet with $\pt$ in the range 50--160\GeV.
Above 160\GeV, the efficiency decreases gradually to
about 30\% for a $\pt$ value of 400\GeV~\cite{Chatrchyan:2012jua}.
All jets are required to satisfy quality selections to minimize the impact of
calorimeter noise and
other sources of misidentified jets~\cite{CMS-PAS-JME-10-003}.
Events are also required to satisfy selection criteria to remove
events with large values of $\ptmiss$ from calorimeter noise,
as described in Ref.~\cite{Chatrchyan:2011tn}.

The jet mass \mjet is calculated from the four-vectors $p_i$ of
all $i$ PF particles clustered into a jet:
\begin{linenomath}
\begin{equation}
\mjet^2 = \Bigl( \sum_{i~\text{in jet}} p_i \Bigr)^2 ,
\end{equation}
\end{linenomath}
where the pion mass is assigned to all charged hadrons.
The reconstruction of \mjet for CA12 jets is studied by using a sample of
highly boosted $\PW \to \PQq \PAQq'$ decays merged into a single jet, as described in Section \ref{sec:uncert}.

\section{Trigger and data}
\label{sec:datasets}
The data were recorded by using single-lepton triggers
with no isolation requirement applied to the leptons.
Events in the muon+jets channel use a trigger that requires
at least one muon with $\pt > 40$\GeV and $\abs{\eta}<2.1$.
The efficiency for this trigger, measured in a $\PZ\to\mu^+\mu^-$ sample,
is 95\% for muons measured within $\abs{\eta}<0.9$,
85\% for muons within $0.9<\abs{\eta}<1.2$, and 83\% for $1.2<\abs{\eta}<2.1$.

The trigger for the electron+jets channel requires at least one electron with $\pt > 30$\GeV in
conjunction with two AK5 jets that have $\pt >100$ and $> 25$\GeV, for the
leading and next-to-leading AK5 jet, respectively. Events are also
included if triggered by a single AK5 jet with $\pt>320\GeV$.
The additional events obtained through this single-jet trigger often contain an
electron merged into a jet that cannot be resolved at the trigger stage.
The resulting combined trigger efficiency is 90\% for events with a leading
AK5 jet with $\pt<320\GeV$. Above this value, the trigger has a turn-on behaviour
and is fully efficient above a value of $350\GeV$.
The trigger efficiencies are measured in data and simulation using a tag-and-probe method
in \Z/$\gamma^*(\to\ell\ell)$+jets and dileptonic \ttbar events.
Small differences between data and simulation are corrected for by
applying scale factors to the simulated events.

{\tolerance=1200
Top quark events, produced via the strong and electroweak interactions, are simulated with
the next-to-leading-order (NLO) generator
\POWHEG 1.380~\cite{Nason:2004rx,Frixione:2007vw,Alioli:2010xd,Alioli:2009je, Re:2010bp}
with a value of $\mt=172.5\GeV$.
The $\PW(\to\ell\nu)$+jets and
$\Z/\gamma^*(\to\ell\ell)$+jets processes are simulated with \MADGRAPH 5.1.5.11 \cite{Alwall:2014hca},
where \MADSPIN \cite{Artoisenet:2012st} is used for the decay of heavy resonances.
Diboson production processes (\PW\PW, \PW\Z, and $\Z\Z$) are simulated
with \PYTHIA~6.424~\cite{pythia6}.
Simulated multijet samples are generated in \MADGRAPH, but constitute
a negligible background.
For the estimation of systematic uncertainties, additional \ttbar samples are
generated with \MCATNLO v3.41~\cite{Frixione:2002ik}
or with \MADGRAPH for seven values of $\mt$
ranging from 166.5 to 178.5\GeV.
\par}

All the samples generated in \MADGRAPH and \POWHEG are interfaced
with \PYTHIA~6 for parton showering and fragmentation
(referred to as \madpyth and \powpyth, respectively).
The MLM algorithm \cite{mlm}
used in \MADGRAPH is applied during the parton matching to avoid double counting
of parton configurations.
The \MADGRAPH samples use the CTEQ6L~\cite{cteq} parton distribution functions (PDFs).
The \POWHEG \ttbar sample uses the CT10~\cite{CT10} PDFs,
whereas the single top quark processes use the CTEQ6M~\cite{Pumplin:2002vw} PDFs.
The \PYTHIA 6 Z2* tune~\cite{Chatrchyan:2013gfi, Khachatryan:2015pea}
is used to model the underlying event.
Top quark events produced with \MCATNLO use the CTEQ6M
PDF set and \HERWIG 6.520~\cite{Corcella:2000bw}
for parton showering and fragmentation (\mcherwig). The default
\HERWIG tune is used to model the underlying event.

The normalisations of the simulated event samples are taken
from the NLO calculations of their cross sections
that contain the next-to-next-to-leading-logarithm (NNLL) soft-gluon resummations for
single top quark production \cite{singletop_xsec},
the next-to-next-to-leading-order (NNLO) calculations for
$\PW(\to\ell\nu)$+jets and $\Z/\gamma^*(\to\ell\ell)$+jets~\cite{fewz1, fewz2, fewz3},
and the NLO calculation for diboson
production~\cite{diboson_xsec}.
The normalisation of the \ttbar simulation is obtained from
QCD NNLO calculations, again
including resummation of NNLL
soft-gluon terms~\cite{Beneke:2011mq, Cacciari:2011hy, Baernreuther:2012ws,
Czakon:2012zr, Czakon:2012pz, Czakon:2013goa, Czakon:2011xx}.

A detailed simulation of particle propagation through the CMS
apparatus and detector response is performed
with \GEANTfour v9.2~\cite{Agostinelli2003250}.
For all simulated samples, the hard collision is
overlaid with simulated minimum-bias collisions. The
resulting events are weighted to reproduce the pileup distribution
measured in data.
The same event reconstruction software is used for data and simulated events.
The resolutions and efficiencies for reconstructed objects are corrected
to match those measured in
data~\cite{Khachatryan:2015hwa, Chatrchyan:2012xi,
Chatrchyan:2012jua, Khachatryan:2014gga, Khachatryan:2016kdb}.

\section{Cross section measurement}
\label{sec:measurement}
\subsection{Strategy}
The measurement is carried out in the $\ell$+jets channel,
which allows the selection of a pure \ttbar sample
because of its distinct signature at large top quark boosts.
The measurement is based on choosing kinematic quantities that do not
bias the \mjet distribution from fully hadronic top quark decays.
A bias would be introduced by, e.g. selecting the leading jet
based on the number of subjets, or requiring a certain maximum value of
the $N$-subjettiness~\cite{Thaler:2010tr, Thaler:2011gf},
as applied in common top quark tagging algorithms~\cite{Kaplan:2008ie, JME-09-001,
htt2009, htt2010, Lapsien:2016zor}. Such a selection would lead to a
distinct three-prong structure of the jet and thus reject events with
one quark being soft or collinear with respect to the momentum of the
top quark decay.

The fiducial region chosen for this investigation is studied through simulations at the particle level
(defined by all particles with lifetimes longer than $10^{-8}$\;s).
The exact selection is detailed below.
It relies on having a highly boosted semileptonic
top quark decay, where the lepton from $\PW \to \ell \nu_\ell$ is close in $\DR$ to the
jet from the hadronisation of the accompanying $\PQb$ quark ($\PQb$ jet).
A second high-\pt jet is selected, which is assumed to originate
from the fully hadronic top quark decay.
A veto on additional jets is employed, which ensures that the
fully hadronic decay is merged into a single jet. The jet veto
is also beneficial for calculating higher-order terms, as it suppresses the
size of nonglobal logarithms~\cite{Chien:2012ur},
which appear because of the sensitivity of the jet mass to radiation
in only a part of the phase space~\cite{Dasgupta:2001sh}.
The event selection at the reconstruction level
is chosen to ensure high efficiency while reducing
non-\ttbar backgrounds.
Finally, the \mjet distribution is unfolded for experimental effects
and then compared to different MC predictions at the particle level.
A measurement of the normalised \mjet distribution is performed as well,
where the normalisation is performed by using the total measured \ttbar cross section
in the fiducial phase-space region.

\subsection{Definition of the fiducial phase space \label{sec:ps}}
The \ttbar cross section as a function of the mass of the leading jet is
unfolded to the particle level, correcting for experimental effects,
with the fiducial phase space at the particle level defined through the
selection described below.

As mentioned previously, the measurement is performed in the $\ell$+jets channel, where
$\ell$ refers to an electron or muon from the $\PW$ boson decay.
The $\tau$ lepton decays are not considered as part of the signal.
Leptons are required to be within $\abs{\eta} < 2.1$ and
have $\pt>45\GeV$.
Jets are clustered by using the CA algorithm with a distance
parameter $R = 1.2$ and required to have $\abs{\eta} < 2.5$.
The value of $R$ is chosen to optimize the relationship between
obtaining a sufficient number of events
and maintaining a narrow width in the jet-mass distribution.
The four-momentum of the leading lepton
is subtracted from the four-momentum of a jet if the lepton is
found within an angular range of ${\Delta R < 1.2}$ of the jet axis.
Events are selected if at least one jet has $\ptone>400\GeV$ and
a second jet has $\pttwo>150\GeV$.
The leading jet in \pt is assumed to originate from the
$\PQt \to \PW\PQb \to \PQq \PAQq^\prime\PQb$ decay, merged into a single jet.
Consequently, the second jet is considered to originate from the fragmented $\PQb$ quark of the
semileptonic top quark decay.
To select events with a highly boosted topology, a veto is employed on
additional jets with $\ptveto>150\GeV$. The jet veto removes about 16\% of the
signal events, but increases the fraction of fully merged top quark decays to
about 40\%, where an event is called fully merged if the maximum distance in $\DR$
between the leading jet at the particle level and each individual parton
from the fully hadronic top quark decay is smaller than $1.2$.

Two additional selection criteria are introduced to ensure that the leading jet
includes all particles from the fully hadronic top quark decay.
The angular difference $\DR(\ell, \text{j}_2)$ between the lepton and the
second jet has to be smaller than 1.2.
This, together with the veto on additional jets, ensures that the top quarks
are produced back-to-back in the transverse plane.
In addition, the invariant mass of the leading jet has to be greater than
the invariant mass of the combination of the second jet and the lepton,
$m_{\text{jet},1} > m_{\text{jet},2+\ell}$.
This improves the choice of the leading jet as originating from the fully hadronic
top quark decay.

\begin{figure}[tb]
\centering
\includegraphics[width=.48\textwidth]{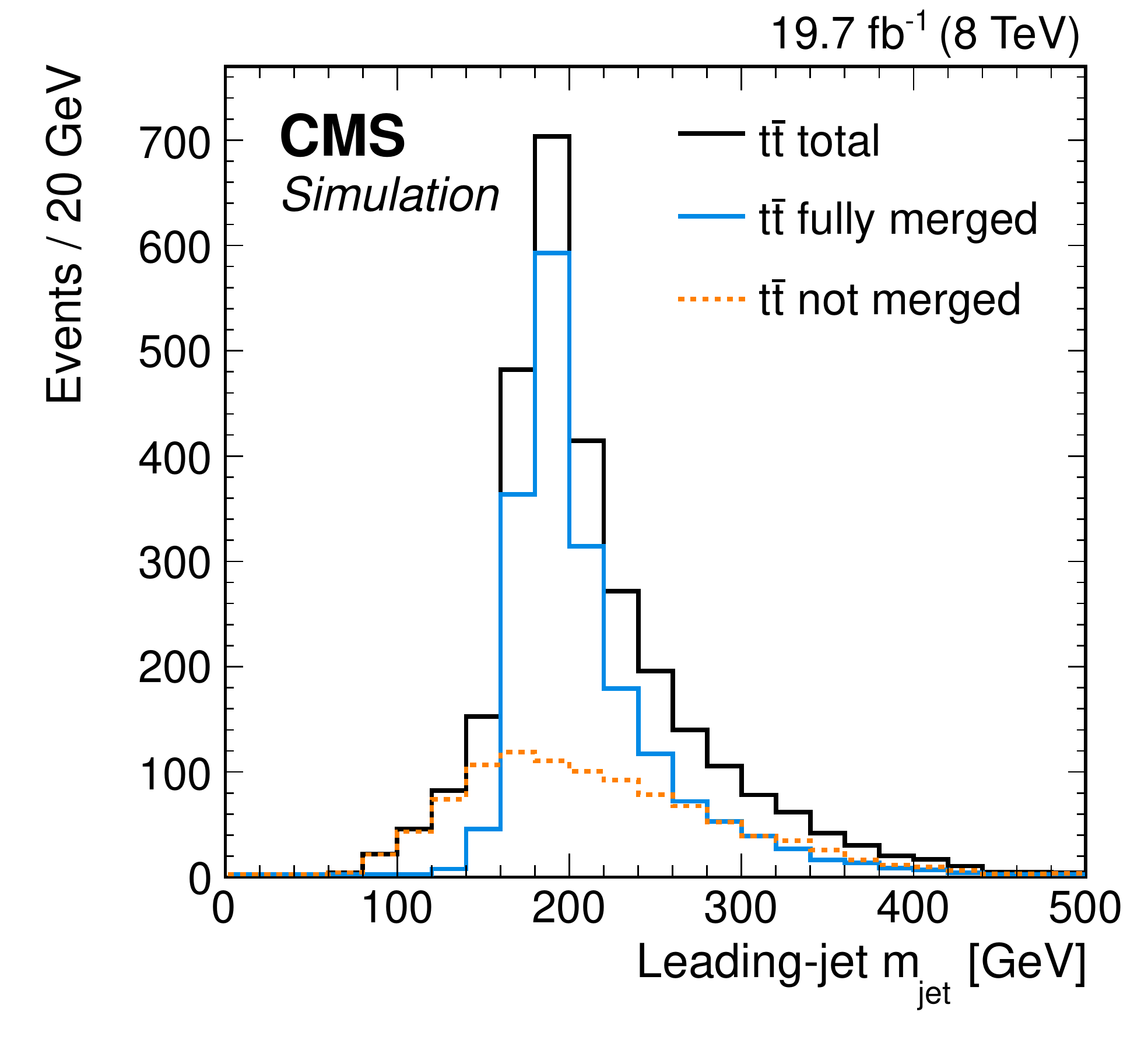}
\caption{Simulated mass distributions of the leading jet in \ttbar events for the $\ell$+jets channel
at the particle level. The events are generated with \powpyth,
and normalised to the integrated luminosity of the data.
The distribution for the total number of selected events (dark solid line) is compared to events
where the leading jet originates from the fully hadronic top quark decay
(light solid line, ``fully merged''), and to events where the leading jet does not
include all the remnants (dotted line, ``not merged'') from the fully hadronic top quark decay.
\label{fig:jetmass_PLev}}
\end{figure}
The simulated distribution of the jet mass at the particle level after this selection
is shown in Fig.~\ref{fig:jetmass_PLev}.
The distribution of all jets passing the particle-level selection is compared to
distributions in jet mass from fully merged and not merged \ttbar decays.
After the selection outlined above, jets that do not originate from
fully merged top quark decays with a fully hadronic final state are expected to constitute
about 35\% of all jets in the final data sample, as determined by using the \powpyth simulation.

\subsection{Selection of events at the reconstruction level}
A selection is applied at the reconstruction level to obtain an enriched \ttbar sample
with high-\pt top quarks, based on leptons without an isolation
requirement.
As a second step, high-\pt jets are required to be
kinematically similar to those selected at the particle level.
Comparable kinematic properties between the reconstruction and particle levels lead
to small bin-to-bin migrations and therefore to small corrections
when unfolding the data.

Selected events must contain exactly one muon or electron with $\pt>45\GeV$ and $\abs{\eta}<2.1$.
Events with more than one lepton are vetoed to suppress
contributions from dileptonic \ttbar decays.
To select highly boosted \ttbar events, at least one AK5 jet
is required to have $\pt>150\GeV$
and another AK5 jet $\pt > 50\GeV$, where both jets
have to fulfil $\abs{\eta}<2.4$.
The suppression of background from multijet production is
accomplished by using a two-dimensional (2D) isolation
variable that is efficient at large top quark boosts, yet notably reduces
multijet background. This 2D isolation requires
the angular difference between the lepton and the nearest
AK5 jet directions $\DR_{\text{min}}(\text{lepton, jets})$ to be greater than $0.5$, or the perpendicular
component of the lepton momentum relative to the nearest AK5 jet
$\ptrel$ to be larger than $25\GeV$.
In the calculation of these quantities, only AK5 jets with $\pt >25\GeV$ are considered.
The efficiency of the 2D isolation requirement
has been studied in data and simulation by using
$\Z/\gamma^*(\to\ell\ell)$+jets events~\cite{Khachatryan:2015sma}.

A requirement on $\ptmiss>20\GeV$ and on the scalar sum $\ptmiss+\pt^\ell > 150\GeV$
reduces the contribution from
multijet and $\Z/\gamma^*(\to\ell\ell)$+jets production,
where $\pt^\ell$ is the lepton transverse momentum.
Given the presence of two $\PQb$ quarks in the events, at least one AK5
jet is required to be identified as originating from the fragmentation of a $\PQb$ quark
by using the CSV algorithm,
which reduces the contribution from $\PW$+jets production.
The electron channel includes an additional topological selection criterion to
suppress the remaining residual contribution from multijet production:
\begin{linenomath}
\begin{equation*}
\abs{ \Delta\phi ( \{ \Pe \,\text{or}\, \text{jet} \} , \,\ptvecmiss ) - 1.5} < \ptmiss / 50\GeV,
\end{equation*}
\end{linenomath}
with $\Delta\phi$ measured in radians and \ptmiss in \GeV.
This criterion rejects events in which $\ptvecmiss$ points along the transverse
momentum vector of the leading jet or the lepton. After these requirements,
the background contribution from multijet production is negligible.

The selection procedure outlined above results in a \ttbar sample with high purity and
selection efficiency at large top quark \pt.
In addition, events are selected with kinematic requirements similar to those at the particle level.
For each event to pass the selection,
at least one jet is required with $\pt>400\GeV$ and another with $\pt > 150\GeV$,
where both jets have to fulfil $\abs{\eta}<2.5$.
Contributions from not fully merged \ttbar events are suppressed with
a veto on additional jets with transverse momentum $\pt > 150\GeV$ and $\abs{\eta} < 2.5$.
The jet veto has an efficiency of $93\%$ for fully-merged signal events.
The fraction of fully merged events with a back-to-back topology
is further enhanced by selecting events with an angular difference
$\DR(\ell, \text{j}_2)<1.2$ between the
directions of the lepton and the subleading jet.
To ensure that the leading jet originates from the fully merged top quark decay,
its invariant mass is required to be larger than the mass of the subleading jet.
With these selection criteria, the reconstruction efficiency for \ttbar events
where one top quark decays semileptonically in the fiducial
region of the measurement is 23.2\%.
Several of the above criteria are relaxed in the unfolding procedure to define
sideband regions included as additional bins in the response matrix, increasing thereby
the reconstruction efficiency.

After the selection procedure, the contribution of non-signal \ttbar events from \ttbar
decays to the $\tau$+jets, dilepton, and all-jets
channels constitute, respectively, $7.3\%$, $11.6\%$, and $0.4\%$ of the
selected events. These contributions are accounted for in the unfolding.

The distributions in $\pt$ and $\eta$ for the leading jet in selected events are shown in
Fig.~\ref{fig:control} from data and simulation.
The mass distribution of the leading jet at the reconstruction level
is shown in Fig.~\ref{fig:mass_reco} for the \pt regions
of $400 < \pt < 500\GeV$ (\cmsLeft) and $\pt > 500\GeV$ (\cmsRight).
In these plots the $\ttbar$ simulation is scaled such that the number of
simulated events matches the number of selected events observed in data.
Overall good agreement between data and the predictions is observed.
The slight slope in the data/MC ratio of the jet mass distribution in
Fig.~\ref{fig:mass_reco} (\cmsLeft) is covered by the jet energy and mass
scale uncertainties, as described below.
\cmsFigPage
  \centering
  \includegraphics[width=.48\textwidth]{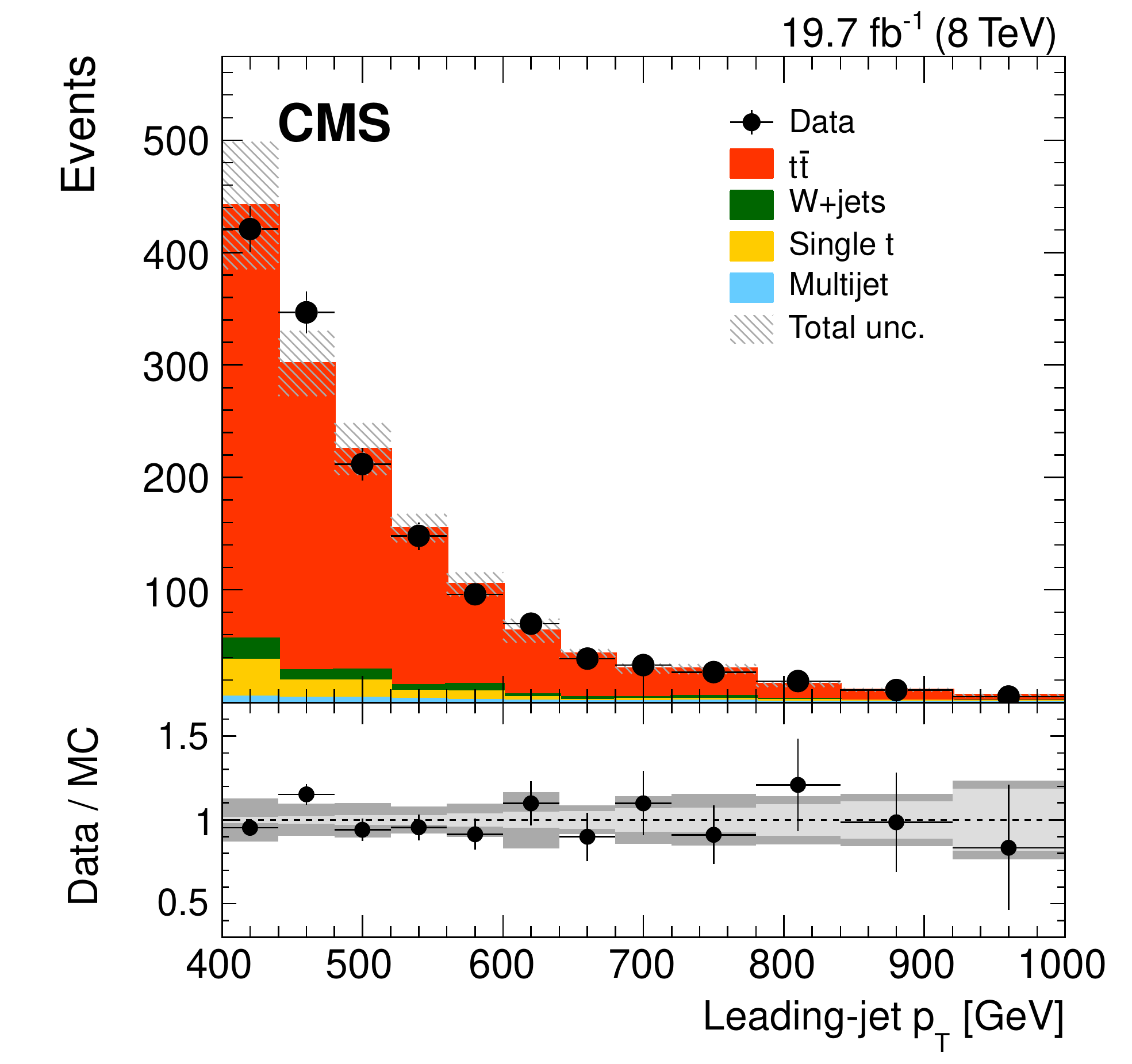}
  \includegraphics[width=.48\textwidth]{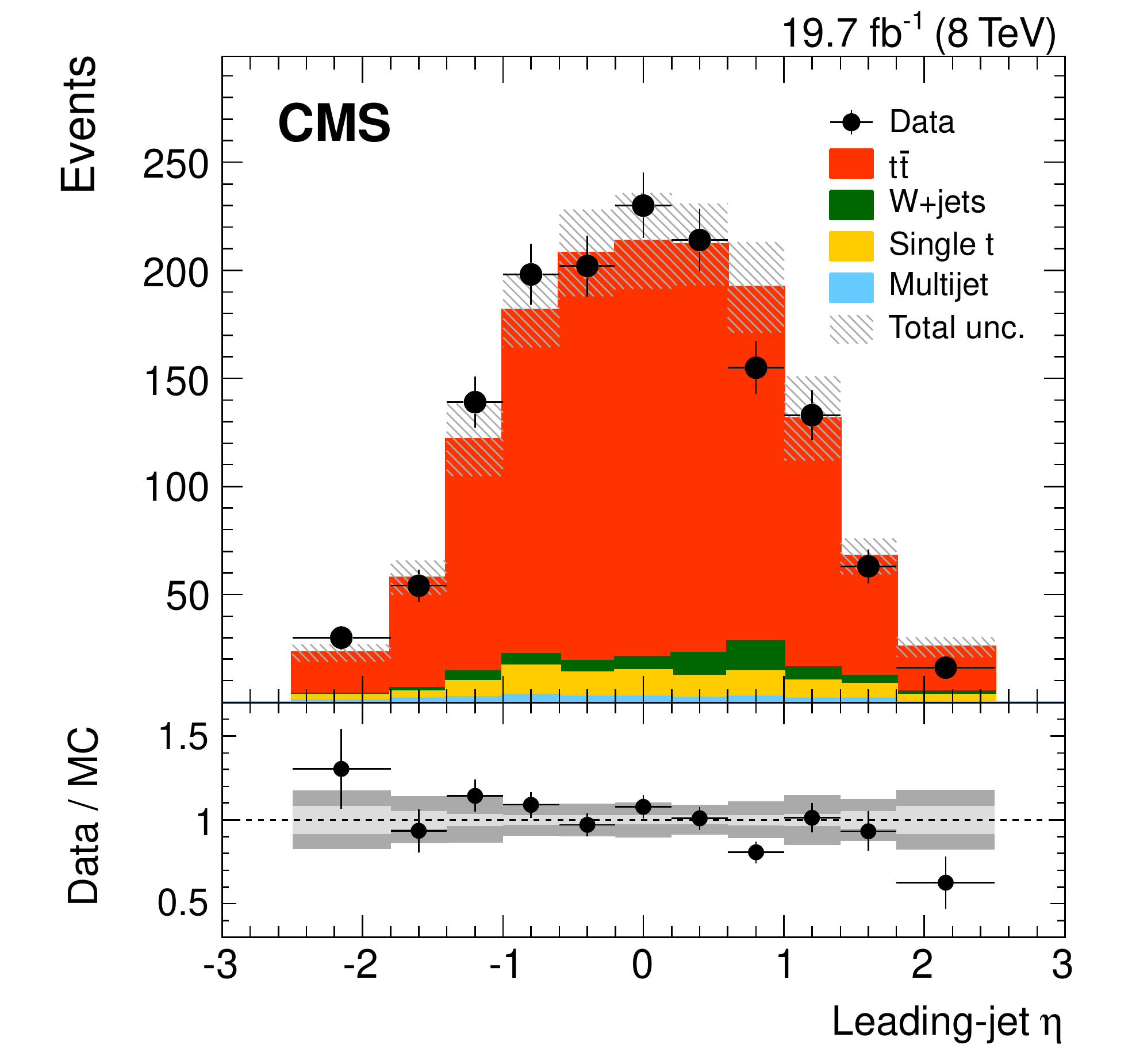}
  \caption{Distributions of $\pt$ (\cmsLeft) and $\eta$ (\cmsRight) of the leading
  jet from data (points) and simulation (filled histograms). The vertical bars on the points
  show the statistical uncertainty and the horizontal bars show the bin widths.
  The electron and muon channels are shown combined.
  The hatched region shows the total uncertainty in the
  simulation, including the statistical and experimental systematic uncertainties. The panels below
  show the ratio of the data to the simulation. The uncertainty bands include the statistical and
  experimental systematic uncertainties, where the statistical (light grey) and total (dark grey)
  uncertainties are shown separately in the ratio.
  \label{fig:control}}
\end{figure}

\cmsFigPage
  \centering
  \includegraphics[width=.48\textwidth]{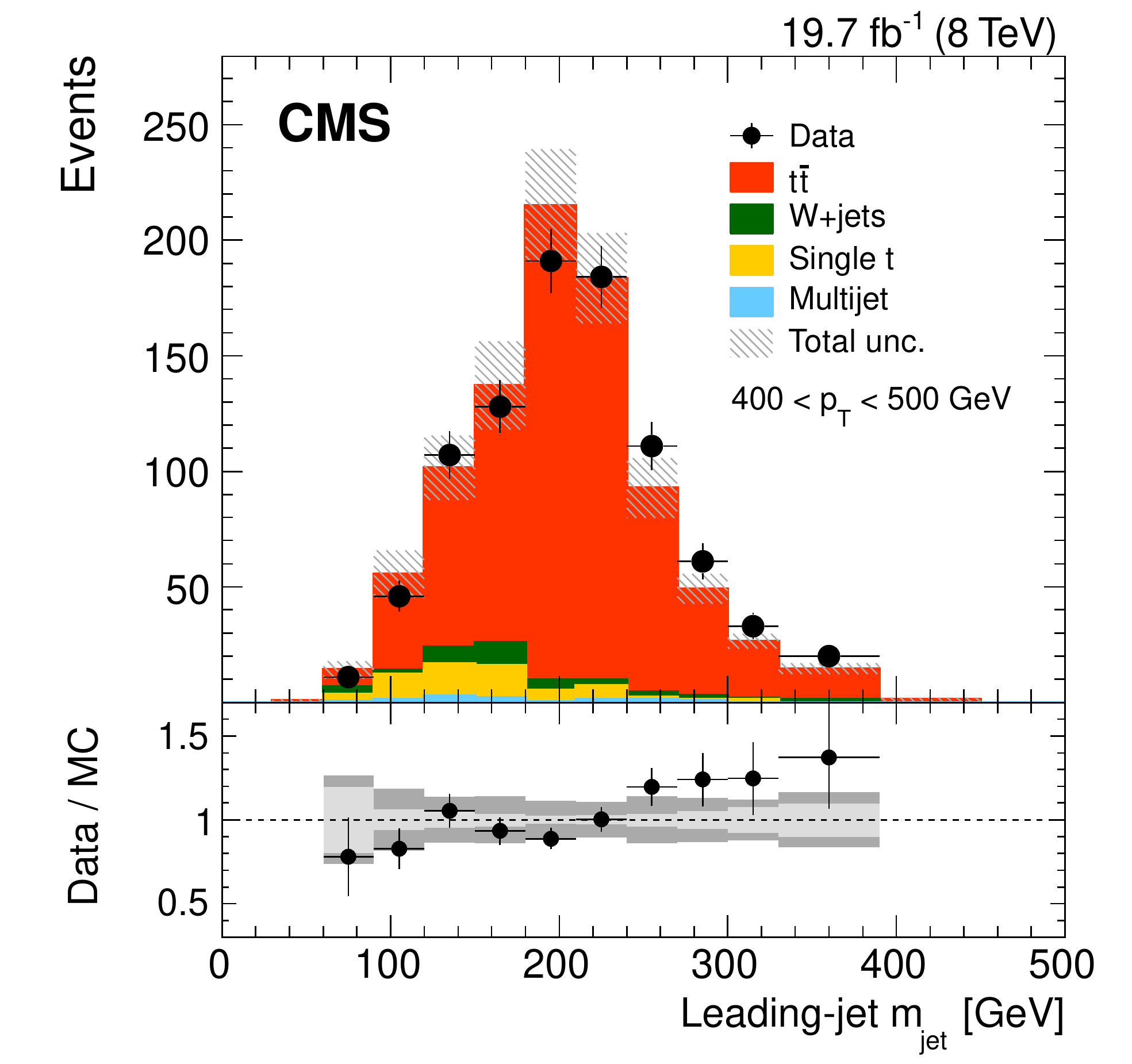}
  \includegraphics[width=.48\textwidth]{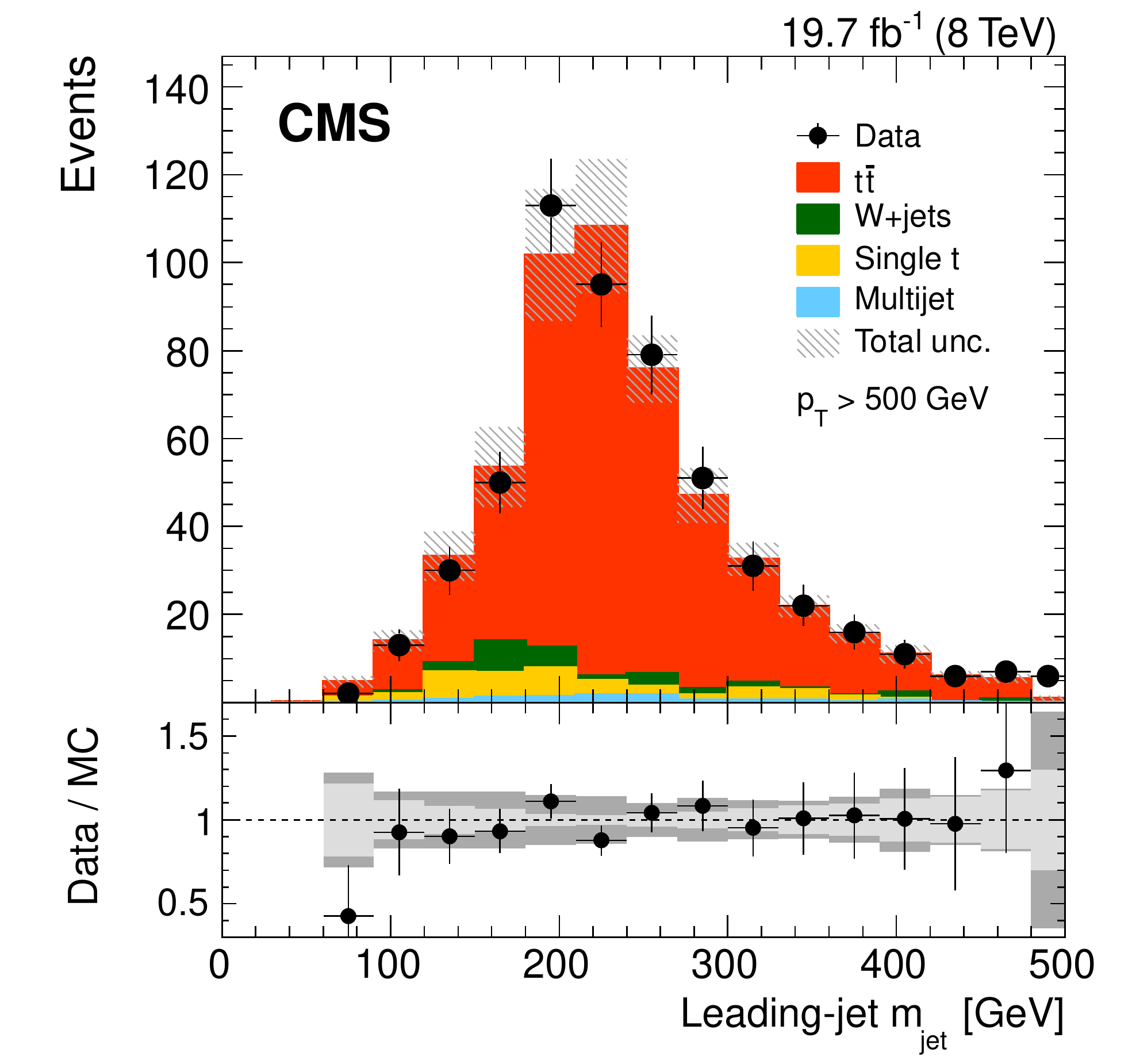}
  \caption{Distributions of the leading-jet invariant mass from data (points)
  and simulation (filled histograms). The vertical bars on the points
  show the statistical uncertainty and the horizontal bars show the bin widths
  for the combined electron and muon channels.
  The distributions for $\pt$ bins of $400 < \pt < 500\GeV$ (\cmsLeft)
  and $\pt > 500\GeV$ (\cmsRight) are given.
  The hatched region shows the total uncertainty in the
  simulation, including the statistical and experimental systematic uncertainties. The panels below
  show the ratio of the data to the simulation. The uncertainty bands include the statistical and
  experimental systematic uncertainties, where the statistical (light grey) and total (dark grey)
  uncertainties are shown separately in the ratio.
  \label{fig:mass_reco}}
\end{figure}

Table~\ref{tab:count} shows the total number of events observed in data
together with the total number of signal and background events determined
from simulation.
\begin{table}[tb]
\centering
\topcaption{Number of events obtained after applying the full selection.
The results are given for the individual sources of background, \ttbar\ signal,
and data. The uncertainties correspond to the statistical and systematic components
added in quadrature.
}
\begin{tabular}{lc}
Source   &  Number of events    \\
\hline
Multijet                &    21 $\pm$   21  \\
$\PW$+jets              &    60 $\pm$   13  \\
Single top quark        &    90 $\pm$   21  \\
Total background        &   171 $\pm$   32  \\[1.5ex]
\ttbar\ signal          &  1410 $\pm$  152  \\[1.5ex]
Data                    &  1434 \\
\end{tabular}
\label{tab:count}
\end{table}

\subsection{Unfolding from the reconstruction level to the particle level}
The transformation from the reconstruction to the particle level is
carried out through a regularised unfolding based on a least-squares
fit, implemented in the {TUnfold}~\cite{Schmitt:2012kp} framework.
This procedure suppresses the statistical fluctuations by a regularisation
with respect to the count in each bin.
The optimal regularisation strength is determined through a
minimization of the average global correlation coefficient of the output
bins~\cite{Schmitt:2016orm}.
Contributions from background processes such as $\PW$+jets, single top quark, and
multijet production are determined from simulation and
subtracted from the data prior to the unfolding.
Non-signal \ttbar events are accounted for in the
unfolding by including them in the response matrix, described below.

The response matrix is evaluated by using $\ttbar$ events simulated with $\powpyth$.
It is obtained for the two regions in the leading-jet \pt of
$400 < \pt < 500\GeV$ and $\pt>500\GeV$. This division is needed
to account for the distribution of the $\pt$ spectrum.
The response matrix includes three additional sideband regions to account for
migrations in and out of the phase-space region of the measurement. These are obtained
for a lower leading-jet $\pt$ of $ 300 < \pt < 400 \GeV$,
a lower second-leading-jet $\pt$ of $100  < \pt < 150 \GeV$,
and a higher veto-jet $\pt$ of $150  < \pt < 200 \GeV$.
Events that are reconstructed, but do not pass the particle-level
selections, are also included in the response matrix.
The electron and muon channels are combined, and the combined distribution is
unfolded to ensure a sufficient number of events in the unfolding procedure.
The electron and muon channels are also unfolded separately, and the
results are compared to verify their consistency.

\subsection{Uncertainties \label{sec:uncert}}

\subsubsection*{Statistical uncertainties}
Statistical uncertainties in the unfolding procedure arise from three sources.
The dominant source reflects the statistical fluctuations in the input data.
Second are the uncertainties from the finite number of simulated events used to
calculate the response matrix. The third source reflects the statistical uncertainties in
the simulation of the background processes.
After the unfolding, a total statistical uncertainty
is obtained for each bin of the \mjet distribution that includes the effects from all three sources,
which are correlated among the individual measurement bins.

\subsubsection*{Experimental systematic uncertainties}
Systematic uncertainties related to experimental effects are evaluated by
changing calibration factors and corrections to efficiencies
within their corresponding uncertainties.
The resulting covariance matrix of the unfolded measurement is
computed through standard error propagation.
The uncertainties are evaluated by unfolding pseudo-data simulated with \madpyth.
Pseudo-data are preferred over data because of the smaller statistical fluctuations
in the estimation of the systematic uncertainties.
The change in each parameter that yields the largest variation in the unfolded measurement is taken
as the uncertainty owing to that parameter. The following sources of experimental systematic
uncertainties are considered.

The applied jet energy corrections (JEC) depend on the $\pt$ and $\eta$ of the
individual jets. The JEC are obtained by using anti-\kt jets with
$R=0.7$ (AK7)~\cite{Khachatryan:2016kdb}, and their use is checked on CA12 jets
by using simulated events.
Residual differences between generated and reconstructed jet momenta
caused by the larger jet size used in this analysis result in
increased uncertainties in the JEC by factors of two to four
with respect to the AK7 values.
Changes of the JEC within their uncertainties are made in the three-momenta of the jets
to estimate the effect on the measured cross section.
The jet mass is kept fixed to avoid double-counting of uncertainties when including
the uncertainty in the jet-mass scale.
A smearing is applied in the jet energy resolution (JER) as an $\eta$-dependent correction
to all jets in the simulation. The corrections are again changed within their uncertainty
to estimate the systematic uncertainty related to the JER smearing. The
uncertainties are found to be small compared to the ones from the JEC.
The jet-mass scale and the corresponding uncertainty in the CA12 jets have been studied in events
that contain a $\PW \to \PQq \PAQq'$ decay reconstructed as a single jet in \ttbar production.
The ratio of the reconstructed jet-mass peak positions in data and simulation is
$1.015 \pm 0.012$.
No correction to the jet-mass scale is applied, but an uncertainty of
$1.5\%$ is assigned, corresponding to the difference in peak positions.
The widths of the jet mass distributions are about 15\GeV, consistent between data
and simulation.

Corrections in $\PQb$ tagging efficiency are applied as $\pt$-dependent
scale factors for each jet flavour. The corresponding systematic uncertainties are
obtained by changing the scale factors within their uncertainties.
Pileup correction factors are applied to match the number of primary interactions
to the instantaneous luminosity profile in data.
The uncertainty is obtained by changing the total inelastic cross section by
${\pm}5\%$~\cite{Antchev:2011vs}.
Trigger and lepton identification scale factors are used to correct for
differences in the lepton selection efficiency between data and simulation.
The corresponding uncertainties are computed by changing
the scale factors within their uncertainties~\cite{Khachatryan:2015hwa, Chatrchyan:2012xi}.

\subsubsection*{Normalisation uncertainties}
The effects from uncertainties in background processes are calculated
by changing the amount of background subtracted prior to the unfolding
and propagating the effect to the output.
The uncertainty in the $\PW$+jets cross section is taken to be 19\%,
as obtained from a measurement of $\PW$+heavy-flavour quark production~\cite{Khachatryan:2016ipq};
an uncertainty of 23\% is applied to the single top quark cross
section~\cite{Chatrchyan:2014tua}; and an uncertainty of 100\% is assumed
for multijet production, estimated from the comparison of various
kinematic distributions between data and simulation.
Uncertainties affecting the overall normalisation are added in quadrature
to the total uncertainty after the unfolding. An uncertainty of 2.6\% is applied
subsequently for the integrated luminosity~\cite{CMS:2013gfa}.

\subsubsection*{Modelling uncertainties}
The unfolding is checked for its dependence on the simulation of \ttbar production through the use
of alternative programs to generate events.
The effect on the measurement is estimated by using one simulation
as pseudo-data input to the unfolding, and another for the calculation of the response matrix.
The unfolded result is then compared to the particle-level distribution from the simulation
used as pseudo-data. Differences between the unfolded result and the truth-level
distribution are taken as the modelling uncertainties.

The uncertainty from the choice of MC generator is estimated by unfolding
pseudo-data simulated with \madpyth through a response matrix evaluated
with \powpyth.
The effect from the choice of the parton-shower simulation is
estimated from events generated with \mcherwig.

The dependence on the choice of $\mtMC$ in the simulation
used to correct the data is also checked.
While the unfolded measurement is largely independent of the choice
of $\mtMC$, residual effects from the kinematic properties of the leptons
and jets can lead to additional uncertainties.
These uncertainties are evaluated by using events simulated with
\MADGRAPH{}+\PYTHIA
for seven values of $\mtMC$ from
166.5 to 178.5\GeV, as pseudo-data.
This range is considered because no measurement of $\mt$ in this
kinematic regime exists, and a stable result, independent of the
specific choice of $\mtMC$, is therefore crucial.
For this check, the response matrix is obtained with \madpyth and a value of
$\mtMC = 172.5\GeV$.
The envelope of the uncertainty obtained for different values of $\mtMC$
is used to define an additional modelling uncertainty.

The uncertainty from the uncalculated higher-order terms in the simulation is estimated
by changing the choice of the factorisation and renormalisation scales $\mu_\mathrm{F}$ and
$\mu_\mathrm{R}$.
For this purpose events simulated with \powpyth are used, where
the scales are changed up and down by factors of two relative to their nominal
value. This is set to $\mu_\mathrm{F}^2 = \mu_\mathrm{R}^2 = Q^2$,
where the scale of the hard process is defined by
$Q^2 = \mt^2 + \sum \pt^2$ with the sum over all additional final-state
partons in the matrix-element calculation.
Events with varied scales are unfolded through a response matrix obtained
with the nominal choice of scales. The uncertainty in the
measurement is defined by the largest change found in the study.

Uncertainties from the PDF are evaluated by using the
eigenvectors of the CT10 PDF set with the \powpyth simulation.
The resulting differences in the response matrix are propagated to the
measurement. The individual uncertainties for each eigenvector
are scaled to the 68\% confidence level and added in quadrature~\cite{CT10}.

\subsubsection*{Summary of uncertainties}

A summary of the relative uncertainties in this measurement
is shown in Fig.~\ref{fig:unc}.
The largest contribution is from the statistical uncertainties.
The experimental systematic uncertainties are far smaller than those from
the modelling of \ttbar production.
The largest uncertainties are expected to improve considerably with
more data at higher centre-of-mass energies.
Besides a reduction of the statistical uncertainties,
an unfolding of the data using finer bins and
as a function of more variables will then be possible,
which will result in a reduction of the
systematic uncertainties from the simulation of \ttbar events. More data will also
allow for a measurement that uses smaller jet sizes, which will reduce the uncertainties
coming from the jet energy and jet mass scales.
\begin{figure}[tbp]
\centering
\includegraphics[width=.48\textwidth]{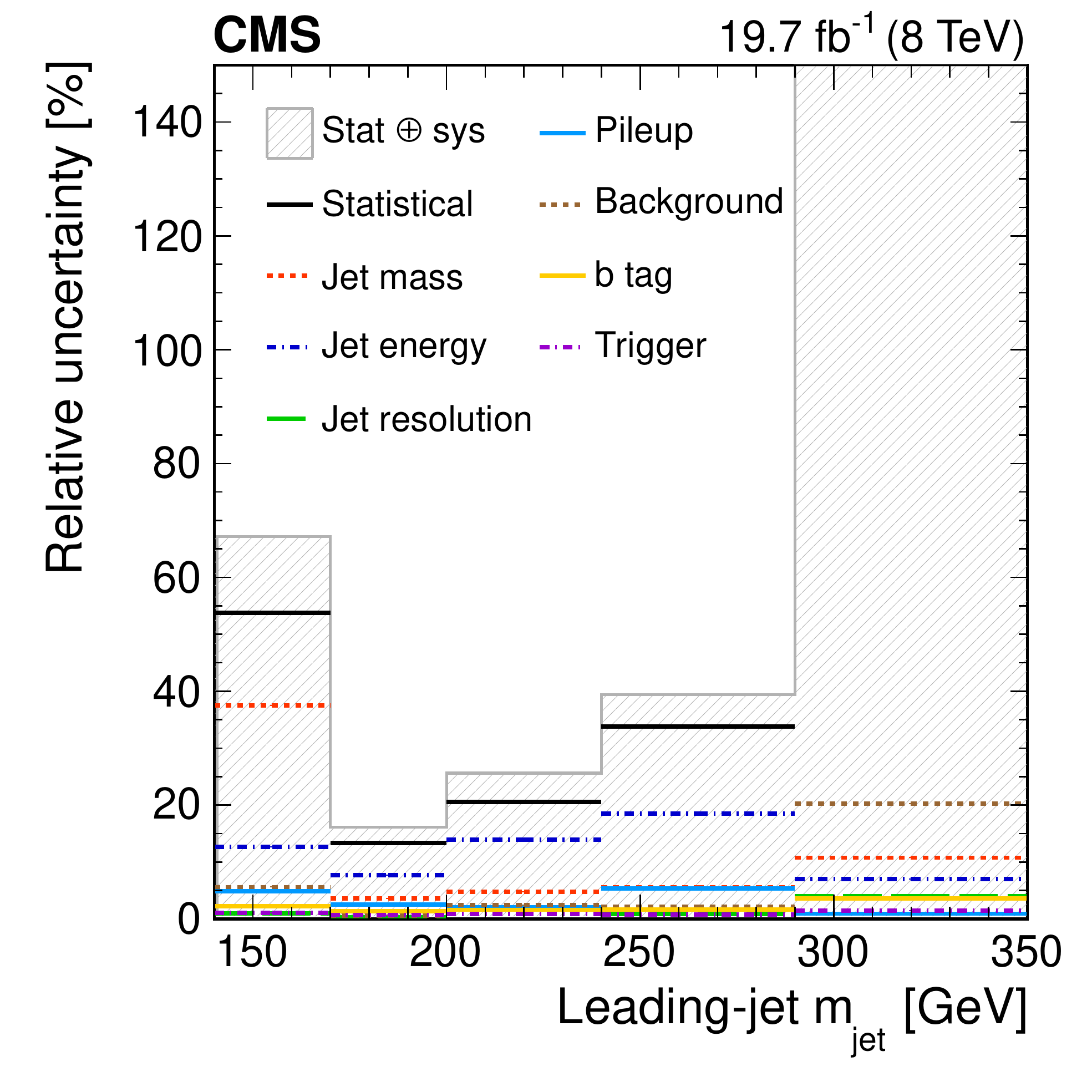}
\includegraphics[width=.48\textwidth]{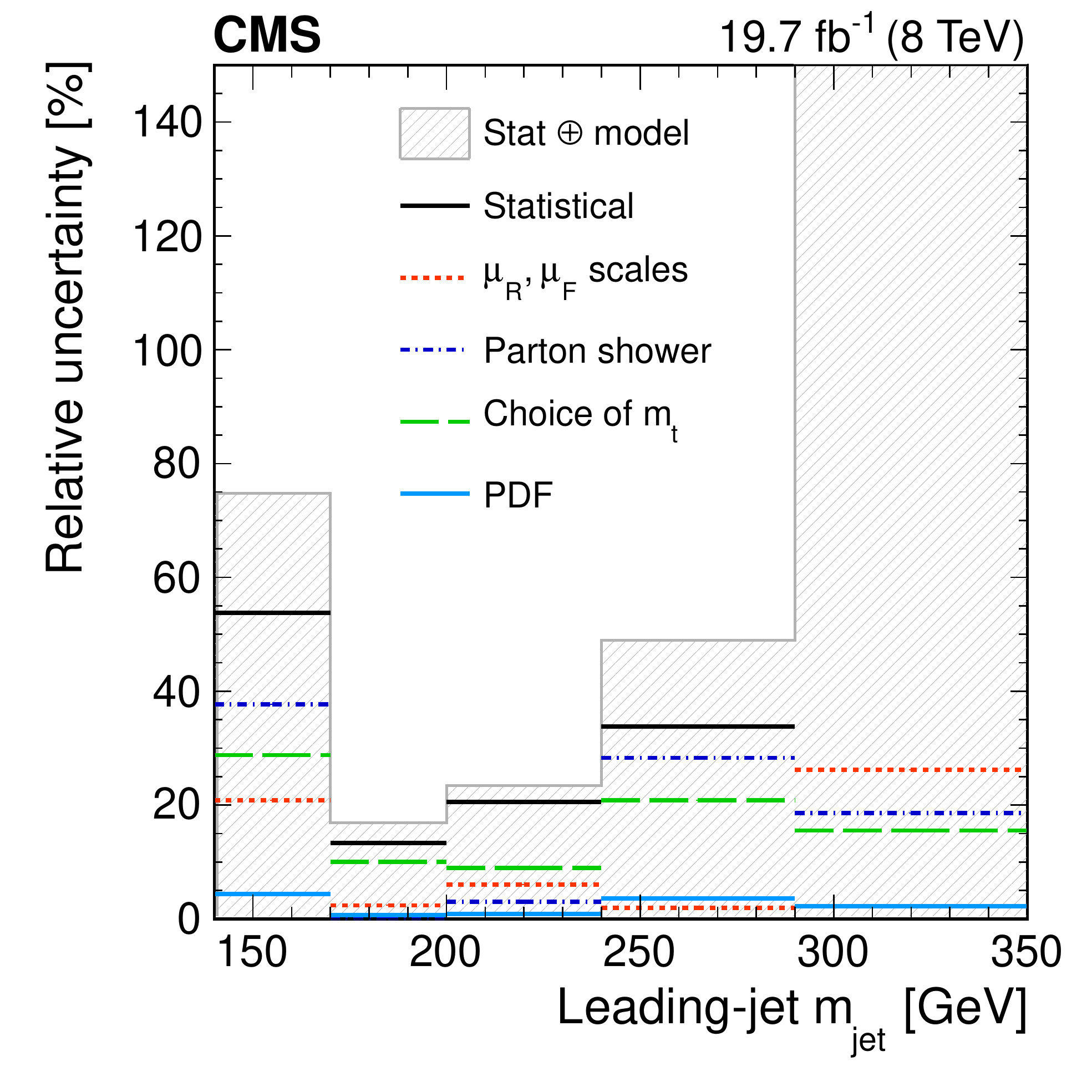}
\caption{Statistical uncertainties compared to the individual experimental systematic
uncertainties (\cmsLeft), and statistical uncertainties compared to
the systematic uncertainties originating from the modelling of \ttbar production (\cmsRight),
as a function of the leading-jet mass.
The total uncertainties are indicated by the grey cross-hatched regions.
The statistical and total uncertainties in the last bin are
around 300\% and exceed the vertical scale.
The size of the horizontal bars represents the bin widths.
\label{fig:unc}}
\end{figure}

\subsection{Cross section results \label{sec:results}}
\begin{table}[bt]
\topcaption{Summary of the selection criteria used to define the fiducial region of the measurement.
\label{tab:fiducial}}
\centering
\begin{tabular}{l | l c l }
Leptons    &   $\pt^\ell>45\GeV$ & & $\abs{\eta^\ell} < 2.1$ \\[0.1cm] \hline
\multirow{3}{*}{Jets}  &   $\ptone>400\GeV$  & \multirow{3}{*}{$\Bigg\}$} & \multirow{3}{*}{$\abs{\eta} < 2.5$} \\
           &   $\pttwo>150\GeV$  &  & \\
           &   $\ptveto>150\GeV$ &  & \\[0.1cm] \hline
\multirow{2}{*}{Event}      &  \multicolumn{3}{l}{$\DR(\ell, \text{j}_2)<1.2$} \\
           &  \multicolumn{3}{l}{$m_{\text{jet},1} > m_{\text{jet},2+\ell}$}
\end{tabular}
\end{table}
The particle-level \ttbar cross section for the fiducial phase-space region
is measured differentially as a function of the leading-jet mass in the
$\ell$+jets channel. The selection criteria defining the fiducial measurement region are
summarised in Table~\ref{tab:fiducial} (cf.\ Section~\ref{sec:ps}).

\begin{figure}[tb]
\centering
\includegraphics[width=.48\textwidth]{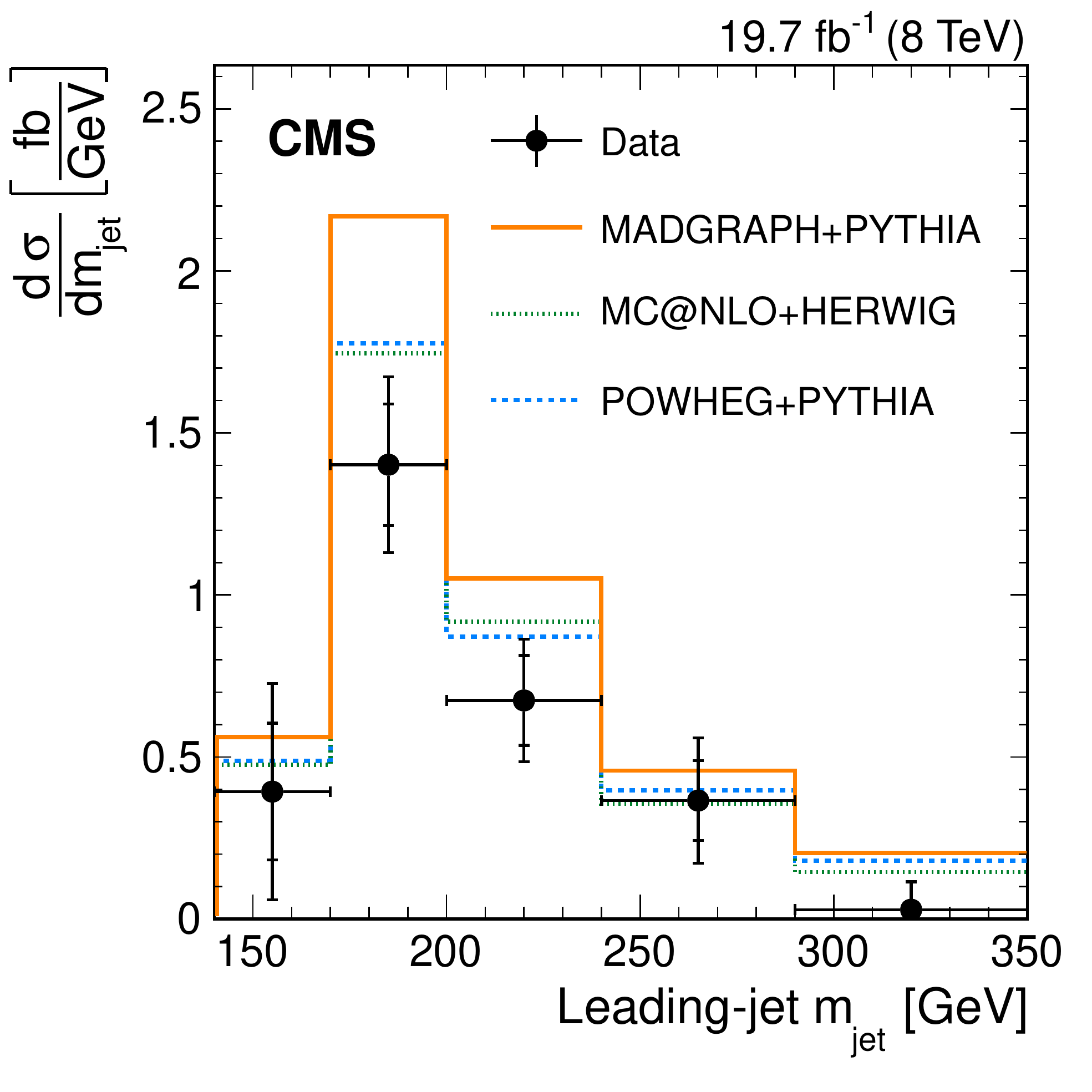}
\caption{Fiducial-region particle-level differential \ttbar cross sections
as a function of the leading-jet mass.
The cross sections from the combined electron and muon channels (points)
are compared to predictions from the \madpyth, \powpyth, and \mcherwig generators (lines).
The vertical bars represent the statistical (inner) and the total (outer) uncertainties.
The horizontal bars show the bin widths.
\label{fig:differential}}
\end{figure}

\begin{table*}
\topcaption{Measured particle-level \ttbar differential cross sections
in the fiducial region as a function of \mjet,
with the individual and total uncertainties in percent.
\label{tab:cross_sections}}
\centering
\begin{tabular}{l | c c c c c }
Range in \mjet [\GeVns{}]            & 140--170  & 170--200  & 200--240  & 240--290  & 290--350 \\ \hline
Integrated cross section [fb]   & 12   & 42   & 27   & 18   & 1.7 \\ \hline
Statistical uncertainty [\%]    & 54   & 13   & 21   & 34   & 300 \\
Systematic uncertainty [\%]     & 40   &  9   & 16   & 20   & 25  \\
Model uncertainty [\%]          & 52   & 10   & 11   & 35   & 36  \\ \hline
Total uncertainty [\%]          & 85   & 19   & 28   & 53   & 300
\end{tabular}
\end{table*}

The measured differential cross section as a function of the leading-jet mass in this
fiducial region is shown in Fig.~\ref{fig:differential},
and the numerical values are given in Table~\ref{tab:cross_sections}.
The full covariance matrices are given in Appendix~\ref{sec:covariance}.
The data are compared to simulated distributions obtained with \powpyth, \madpyth, and \mcherwig.
The total measured \ttbar cross section for $140 < \mjet < 350\GeV$
in the fiducial region is
$\sigma = 101 \pm 11\stat\pm 13\syst \pm 9\,(\text{model})\fb$,
where the last uncertainty is from the modelling of the \ttbar signal.
Combining all the uncertainties in quadrature gives a value of $\sigma = 101 \pm 19\fb$.
The predicted fiducial-region cross sections from the \madpyth and \powpyth \ttbar simulations,
assuming a total \ttbar cross section of 253\pb~\cite{Beneke:2011mq, Cacciari:2011hy,
Baernreuther:2012ws, Czakon:2012zr, Czakon:2012pz, Czakon:2013goa, Czakon:2011xx},
are $159\,^{+17}_{-18}$ and $133\,^{ +18}_{ -28}\fb$, respectively,
where the uncertainties are systematic and come from the variations of
$\mu_\mathrm{R}$ and $\mu_\mathrm{F}$.
The predictions exceed the measurements, consistent with previously measured \ttbar
cross sections at large top quark \pt~\cite{Aad:2015hna, Khachatryan:2016gxp}.
A similar trend is observed when comparing the data to the prediction from \mcherwig.
Recent NNLO calculations~\cite{Czakon:2015owf} of the top quark \pt spectrum alleviate
this discrepancy.

{\sloppy
The normalised differential cross section $(1/\sigma) (\rd\sigma/\rd\mjet{})$
is obtained by dividing the differential cross sections by
the total cross section in the \mjet range from 140 to 350\GeV.
The result is shown in Fig.~\ref{fig:normalized}, together with the predictions of
\madpyth for three values of $\mtMC$.
The numerical values of the measured particle-level cross sections are given in
Table~\ref{tab:cross_sections_norm}, together with the individual and total uncertainties.
The covariance matrices of the measurement are given in Appendix~\ref{sec:covariance}.
The data are well described by the simulation, showing that the overall
modelling of the top quark jet mass is acceptable, once the disagreement with
the total cross section at large \pt is eliminated by the normalisation.
The sensitivity of the measurement to \mt is clearly visible,
albeit compromised by the overall uncertainties.
\par}

\begin{figure}[tb]
\centering
\includegraphics[width=0.48\textwidth]{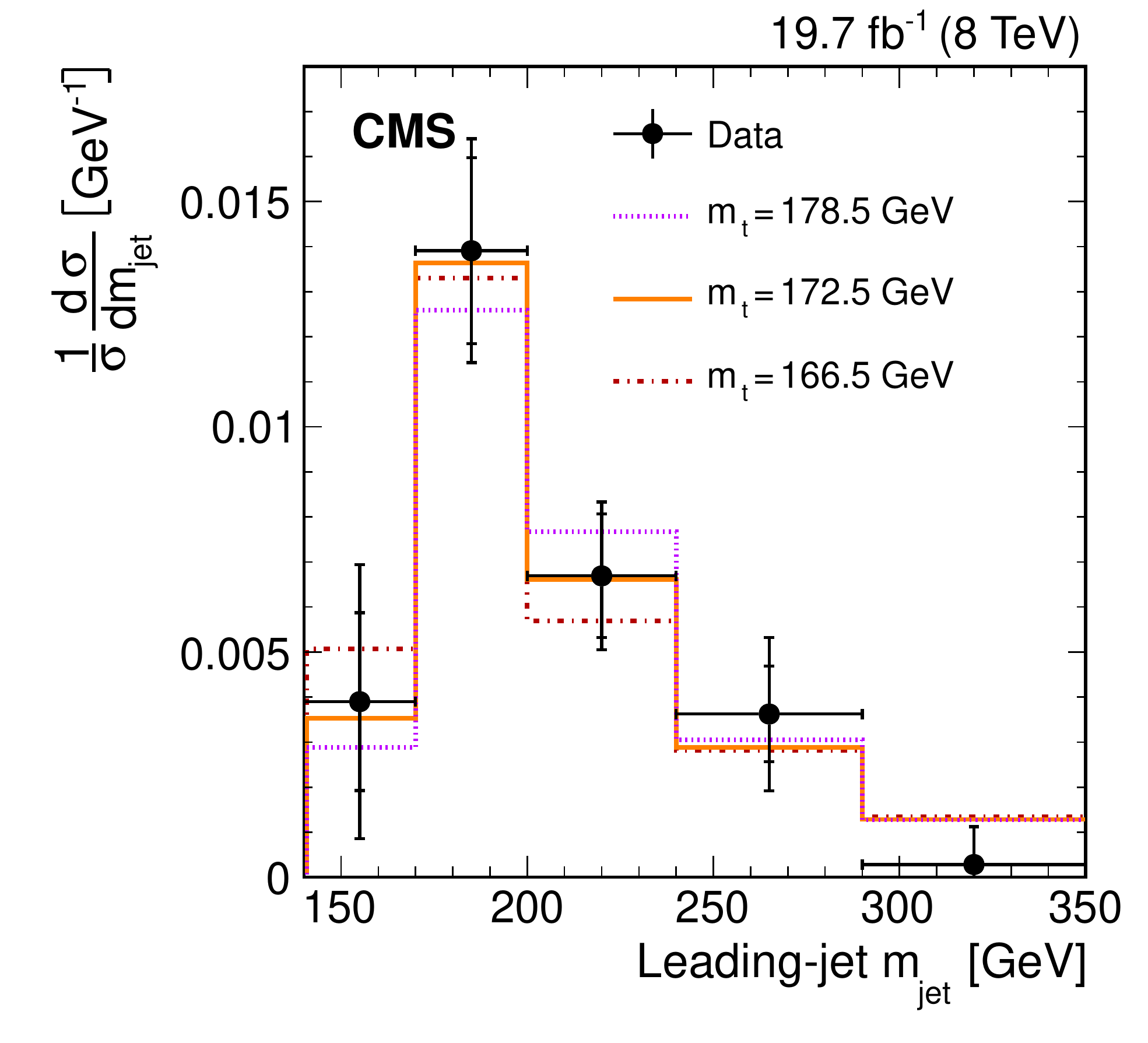}
\caption{The normalised particle-level \ttbar differential cross section
in the fiducial region as a function of the leading-jet mass.
The measurement is compared to predictions from \madpyth for three
values of \mtMC.
The vertical bars represent the statistical (inner) and the total (outer) uncertainties.
The horizontal bars show the bin widths.
\label{fig:normalized}}
\end{figure}
\begin{table*}[tb]
\topcaption{Values of the particle-level \ttbar
differential cross section in the fiducial region, normalized
to unity, as a function of the leading-jet mass.
The individual and total uncertainties are given in percent.
\label{tab:cross_sections_norm}}
\centering
\begin{tabular}{l | c c c c c }
Range in \mjet [\GeVns{}]                 & 140--170  & 170--200  & 200--240  & 240--290  & 290--350 \\ \hline
Integrated normalised cross section  & 0.12  & 0.42 & 0.27 & 0.18 & 0.017 \\ \hline
Statistical uncertainty [\%]         & 51    & 15   & 21   & 29   & 290   \\
Systematic uncertainty [\%]          & 34    &  5   &  9   & 13   & 27    \\
Model uncertainty [\%]               & 48    &  9   & 10   & 34   & 36    \\ \hline
Total uncertainty [\%]               & 78    & 18   & 25   & 47   & 300
\end{tabular}
\end{table*}

\section{Sensitivity to the top quark mass}
\label{sec:top_mass}
Calculations of \mjet for \ttbar production from first principles, by using a
well-defined definition of \mt and not relying on parton shower and
hadronisation models, are not yet available for the LHC. Still, a determination of the
top quark mass parameter in general-purpose event generators that uses
the normalised particle-level cross sections provides a
proof of principle for the feasibility of the method, a cross-check on
other determinations of \mt, and an estimate of the current measurement's sensitivity.
The value of \mtMC is determined from the normalised differential cross section measurements given in
Table~\ref{tab:cross_sections_norm}, since only the shape of the \mjet distribution can be reliably calculated. Correlations are taken into account through the
full covariance matrix of the measurement, which is given in Appendix~\ref{sec:covariance}.
Theoretical predictions are obtained from
\madpyth for different values of \mtMC.
A fit is performed based on the $\chi^2$ evaluated as
$\chi^2 = d^T V^{-1} d$,
where $d$ is the vector of differences between the measured normalised cross sections
and the predictions, and $V$ is the covariance matrix, which includes the
statistical, experimental systematic, modelling, and theoretical uncertainties.
The latter are calculated by changing up and down by factors of two the scales
$\mu_\mathrm{R}$ and $\mu_\mathrm{F}$ in the \madpyth simulation.
The resulting uncertainties are added to the covariance matrix.
The $\chi^2$ values obtained for different values of \mtMC are fitted by a second-order polynomial
to determine the minimum, and the uncertainty is determined by a change in $\chi^2$ of 1.0.
The result is
\ifthenelse{\boolean{cms@external}}{
\begin{linenomath}
\begin{align}
\mtMC =& 170.8 \pm 6.0\stat \pm 2.8\syst\label{top_mass_extr_Data_individual}\\
&\pm 4.6\,\text{(model)} \pm 4.0\thy\GeV \notag\\
      =& 170.8 \pm 9.0\GeV, \label{top_mass_extr_Data_total}
\end{align}
\end{linenomath}
    }{
\begin{linenomath}
\begin{align}
\mtMC =& 170.8 \pm 6.0\stat \pm 2.8\syst\pm 4.6\,\text{(model)} \pm 4.0\thy\GeV\label{top_mass_extr_Data_individual} \\
      =& 170.8 \pm 9.0\GeV,\label{top_mass_extr_Data_total}
\end{align}
\end{linenomath}
}
where the total uncertainty in Eq.~\eqref{top_mass_extr_Data_total} is the sum in quadrature of
the individual uncertainties in Eq.~\eqref{top_mass_extr_Data_individual}. The fit has a minimum
$\chi^2$ of $1.6$ for three degrees of freedom.
This measurement is the first
determination of \mtMC from boosted \ttbar production, calibrated to the \madpyth simulation.
It is consistent with recent determinations of
\mtMC that use MC event generators~\cite{Aad:2015nba, Khachatryan:2015hba,
Khachatryan:2016wqo, Aaboud:2016igd},
cross section measurements~\cite{Aad:2014kva, Aad:2015waa, Khachatryan:2016mqs},
and indirect constraints from electroweak fits~\cite{Baak:2014ora}.

\section{Summary and outlook}
\label{sec:summary}
The first measurement of the differential \ttbar cross section has been performed
in the $\ell$+jets channel as a function
of the leading-jet mass \mjet in the highly boosted top quark regime.
The measurement is carried out in a fiducial region with fully merged
top quark decays in hadronic final states,
corrected to the particle level.
The normalised differential cross section as a function of \mjet agrees with predictions
from simulations, indicating the good quality of modelling the jet mass in highly boosted top quark decays.
The total fiducial-region cross section for \mjet between 140 and 350\GeV is
measured to be $101 \pm 19\fb$, which is below the predicted value. This
difference is consistent with earlier measurements of a
softer top quark \pt spectrum observed in data than in
simulation~\cite{Aad:2015hna, Khachatryan:2016gxp}.
This measurement is a first step towards measuring unfolded jet substructure distributions
in highly boosted top quark decays. A detailed understanding of
these is crucial for measurements and searches for new physics
making use of top quark tagging algorithms.

The peak position in the \mjet distribution is sensitive to the
top quark mass \mt. This can be used for an independent determination of \mt
in the boosted regime, with the prospect of reaching a more reliable correspondence
between the top quark mass in any well-defined renormalisation scheme
and the top quark mass parameter in general-purpose event generators.

The normalised particle-level \ttbar differential cross section measurement
as a function of \mjet is used to extract a
value of \mtMC in order to estimate the current sensitivity of the data.
The value obtained, $\mtMC = 170.8 \pm 9.0\GeV$,
is consistent with the current LHC and Tevatron average of
$173.34 \pm 0.27\stat\pm 0.71\syst\GeV$~\cite{ATLAS:2014wva},
albeit with a much larger uncertainty.

New data at higher centre-of-mass energies
and with larger integrated luminosities will lead to an improvement in the statistical
uncertainty. More data can also lead to reductions in the experimental
systematic uncertainties, most notably that from the jet mass scale, which is expected to
improve with smaller jet distance parameters.
In addition, improvements in the modelling uncertainty are expected
because of stronger constraints on the simulation in the highly boosted regime.
A reduction in the theoretical uncertainty is also foreseen with the emergence of
higher-order calculations.
The results obtained in this analysis show the feasibility of the method
to obtain the top quark mass in the highly boosted regime.
This can provide an important ingredient for studies
of the relation between the value of the top quark mass
obtained from MC event generators and the
one obtained from first-principle calculations.

\begin{acknowledgments}
We congratulate our colleagues in the CERN accelerator departments for the excellent performance of the LHC and thank the technical and administrative staffs at CERN and at other CMS institutes for their contributions to the success of the CMS effort. In addition, we gratefully acknowledge the computing centres and personnel of the Worldwide LHC Computing Grid for delivering so effectively the computing infrastructure essential to our analyses. Finally, we acknowledge the enduring support for the construction and operation of the LHC and the CMS detector provided by the following funding agencies: BMWFW and FWF (Austria); FNRS and FWO (Belgium); CNPq, CAPES, FAPERJ, and FAPESP (Brazil); MES (Bulgaria); CERN; CAS, MoST, and NSFC (China); COLCIENCIAS (Colombia); MSES and CSF (Croatia); RPF (Cyprus); SENESCYT (Ecuador); MoER, ERC IUT, and ERDF (Estonia); Academy of Finland, MEC, and HIP (Finland); CEA and CNRS/IN2P3 (France); BMBF, DFG, and HGF (Germany); GSRT (Greece); OTKA and NIH (Hungary); DAE and DST (India); IPM (Iran); SFI (Ireland); INFN (Italy); MSIP and NRF (Republic of Korea); LAS (Lithuania); MOE and UM (Malaysia); BUAP, CINVESTAV, CONACYT, LNS, SEP, and UASLP-FAI (Mexico); MBIE (New Zealand); PAEC (Pakistan); MSHE and NSC (Poland); FCT (Portugal); JINR (Dubna); MON, RosAtom, RAS, RFBR and RAEP (Russia); MESTD (Serbia); SEIDI, CPAN, PCTI and FEDER (Spain); Swiss Funding Agencies (Switzerland); MST (Taipei); ThEPCenter, IPST, STAR, and NSTDA (Thailand); TUBITAK and TAEK (Turkey); NASU and SFFR (Ukraine); STFC (United Kingdom); DOE and NSF (USA).

\hyphenation{Rachada-pisek} Individuals have received support from the Marie-Curie programme and the European Research Council and EPLANET (European Union); the Leventis Foundation; the A. P. Sloan Foundation; the Alexander von Humboldt Foundation; the Belgian Federal Science Policy Office; the Fonds pour la Formation \`a la Recherche dans l'Industrie et dans l'Agriculture (FRIA-Belgium); the Agentschap voor Innovatie door Wetenschap en Technologie (IWT-Belgium); the Ministry of Education, Youth and Sports (MEYS) of the Czech Republic; the Council of Science and Industrial Research, India; the HOMING PLUS programme of the Foundation for Polish Science, cofinanced from European Union, Regional Development Fund, the Mobility Plus programme of the Ministry of Science and Higher Education, the National Science Center (Poland), contracts Harmonia 2014/14/M/ST2/00428, Opus 2014/13/B/ST2/02543, 2014/15/B/ST2/03998, and 2015/19/B/ST2/02861, Sonata-bis 2012/07/E/ST2/01406; the National Priorities Research Program by Qatar National Research Fund; the Programa Clar\'in-COFUND del Principado de Asturias; the Thalis and Aristeia programmes cofinanced by EU-ESF and the Greek NSRF; the Rachadapisek Sompot Fund for Postdoctoral Fellowship, Chulalongkorn University and the Chulalongkorn Academic into Its 2nd Century Project Advancement Project (Thailand); and the Welch Foundation, contract C-1845.
\end{acknowledgments}

\appendix
\section{Covariance matrices}
\label{sec:covariance}
The covariance matrices that involve just the statistical components, and the ones
involving the total uncertainty (\ie the statistical, experimental systematic, and
modelling uncertainties) are provided in this appendix.
All experimental, as well as the PDF and parton-shower uncertainties,
are treated as fully correlated in the calculation of the covariance matrices.
The uncertainties in the renormalisation and factorisation scale include correlations
in the first three bins, and the uncertainties coming from the choice of \mtMC
are treated as uncorrelated.
Bins 1 to 5 correspond to the following ranges in \mjet:
140--170, 170--200, 200--240, 240--290, and 290--350\GeV.
The covariance matrices for the differential \mjet measurement
are given in Tables~\ref{tab:covariance_stat} and \ref{tab:covariance}
for the statistical and total uncertainties, respectively.
The covariance matrices for the normalised measurement are given in
Tables~\ref{tab:covariance_stat_norm} and \ref{tab:covariance_norm}.
Note that the covariance matrices of the normalised measurement are singular,
and only four out of the five measurement bins are used in the determination of \mtMC.
\begin{table}[htbp]
\topcaption{Covariance matrix for the statistical uncertainties in the differential
cross section.
All entries are given in units of [fb$^2$]. }
\centering
\cmsTableResize{\begin{tabular}{c|*{5}{.} }
  Bin                       & \multicolumn{1}{c}{1}        & \multicolumn{1}{c}{2}          &\multicolumn{1}{c}{3}          & \multicolumn{1}{c}{4}         & \multicolumn{1}{c}{5}           \\ \hline
1 &  +40.1    &   -4.3    &   -8.0    &   -0.2    &  -0.6    \\
2 &           &  +31.7    &   -1.5    &   -8.1    &  +0.8    \\
3 &           &           &  +30.7    &   +1.0    &  -4.5    \\
4 &           &           &           &  +38.1    &  +7.3    \\
5 &           &           &           &           &  +26.2   \\
\end{tabular}}
\label{tab:covariance_stat}
\end{table}
\begin{table}[htbp]
\topcaption{Covariance matrix for the total uncertainties in the
differential cross section,
including all systematic and modelling uncertainties.
All entries are given in units of [fb$^2$]. }
\centering
\cmsTableResize{\begin{tabular}{c|x*{4}{.} }
  Bin                       & \multicolumn{1}{c}{1}        & \multicolumn{1}{c}{2}          &\multicolumn{1}{c}{3}          & \multicolumn{1}{c}{4}         & \multicolumn{1}{c}{5}           \\ \hline
1 &  +100.4    &  +10.4    &   -0.3    &  -22.5    &  +1.6    \\
2 &            &  +66.1    &  +11.0    &   +1.4    &  +0.8    \\
3 &            &           &  +57.4    &  +12.0    &  -4.7    \\
4 &            &           &           &  +93.8    &  +5.3    \\
5 &            &           &           &           &  +26.7   \\
\end{tabular}}
\label{tab:covariance}
\end{table}
\begin{table}[htbp]
\topcaption{Covariance matrix for the statistical uncertainties in the normalised
differential cross section.
All entries are given in units of [$10^{-4}$]. }
\centering
\cmsTableResize{\begin{tabular}{c|*{5}{.} }
  Bin                       & \multicolumn{1}{c}{1}        & \multicolumn{1}{c}{2}          &\multicolumn{1}{c}{3}          & \multicolumn{1}{c}{4}         & \multicolumn{1}{c}{5}           \\ \hline
1 &  +35.0    &  -11.2    &  -13.0    &   -6.7    &  -4.2    \\
2 &           &  +38.3    &   +0.7    &  -17.2    &  -10.6   \\
3 &           &           &  +30.1    &   -6.0    &  -11.8   \\
4 &           &           &           &  +28.1    &  +1.8    \\
5 &           &           &           &           &  +24.8   \\
\end{tabular}}
\label{tab:covariance_stat_norm}
\end{table}
\begin{table}[htbp]
\topcaption{Covariance matrix for the total uncertainties in the normalised
differential cross section,
including all systematic and modelling uncertainties.
All entries are given in units of [$10^{-4}$]. }
\centering
\cmsTableResize{\begin{tabular}{c|*{5}{.} }
  Bin                       & \multicolumn{1}{c}{1}        & \multicolumn{1}{c}{2}          &\multicolumn{1}{c}{3}          & \multicolumn{1}{c}{4}         & \multicolumn{1}{c}{5}           \\ \hline
1 &  +83.2    &  -18.9    &  -21.0    &  -40.7    &  -2.6     \\
2 &           &  +55.5    &   -2.6    &  -23.7    &  -10.4    \\
3 &           &           &  +43.1    &   -7.4    &  -12.0    \\
4 &           &           &           &  +72.4    &  -0.5     \\
5 &           &           &           &           &  +25.4    \\
\end{tabular}}
\label{tab:covariance_norm}
\end{table}
\clearpage

\bibliography{auto_generated}

\cleardoublepage \section{The CMS Collaboration \label{app:collab}}\begin{sloppypar}\hyphenpenalty=5000\widowpenalty=500\clubpenalty=5000\textbf{Yerevan Physics Institute,  Yerevan,  Armenia}\\*[0pt]
A.M.~Sirunyan, A.~Tumasyan
\vskip\cmsinstskip
\textbf{Institut f\"{u}r Hochenergiephysik,  Wien,  Austria}\\*[0pt]
W.~Adam, E.~Asilar, T.~Bergauer, J.~Brandstetter, E.~Brondolin, M.~Dragicevic, J.~Er\"{o}, M.~Flechl, M.~Friedl, R.~Fr\"{u}hwirth\cmsAuthorMark{1}, V.M.~Ghete, C.~Hartl, N.~H\"{o}rmann, J.~Hrubec, M.~Jeitler\cmsAuthorMark{1}, A.~K\"{o}nig, I.~Kr\"{a}tschmer, D.~Liko, T.~Matsushita, I.~Mikulec, D.~Rabady, N.~Rad, B.~Rahbaran, H.~Rohringer, J.~Schieck\cmsAuthorMark{1}, J.~Strauss, W.~Waltenberger, C.-E.~Wulz\cmsAuthorMark{1}
% ...
\vskip\cmsinstskip
% ...
1:~~Also at Vienna University of Technology, Vienna, Austria\\
% ...

\end{sloppypar}
\end{document}